\documentclass[aps,prb,twocolumn,amsmath,amssymb,floatfix,
superscriptaddress]{revtex4}
\usepackage{tabularx}
\usepackage{bm}
\usepackage{euscript}
\usepackage{epsfig,psfrag,subfigure}
\usepackage{graphicx}
\usepackage{color}
\usepackage{amsfonts}
\usepackage{exscale}
\usepackage{amsbsy}

\begin{document}
\title{Myriad phases of the Checkerboard Hubbard Model}
\author{Hong Yao}
\affiliation{Department of Physics, Stanford University, Stanford, CA 94305}
\author{Wei-Feng Tsai}
\affiliation{Department of Physics, Stanford University, Stanford, CA 94305}
\affiliation{Department of Physics and Astronomy, University of California, Los Angeles, CA 90095}
\author{Steven A. Kivelson}
\affiliation{Department of Physics, Stanford University, Stanford, CA 94305}

\date{\today}
\newcommand{\br}{\mathbf{r}}
\newcommand{\brprime}{{\mathbf{r}^\prime}}
\newcommand{\bR}{\mathbf{R}}
\newcommand{\bRprime}{{\mathbf{R}^\prime}}
\newcommand{\bk}{\mathbf{k}}
\newcommand{\bp}{\mathbf{p}}
\newcommand{\bq}{\mathbf{q}}
\newcommand{\bS}{\mathbf{S}}
\newcommand{\te}{\mathrm{e}}
\newcommand{\btau}{\boldsymbol{\tau}}
\newcommand{\eff}{\mathrm{eff}}
\newcommand{\Tc}{\mathrm{T}_\mathrm{c}}
\newcommand{\bra}{\boldsymbol{\langle}}
\newcommand{\ket}{\boldsymbol{\rangle}}

\begin{abstract}
The zero-temperature phase diagram of the checkerboard Hubbard model is obtained in the {\it solvable} limit in which it consists of weakly coupled square plaquettes. As a function of the on-site Coulomb repulsion $U$ and the density of holes per site, $x$, we demonstrate the existence of {\em at least} 16 distinct phases. For instance, at zero doping, the ground state is a novel $d$-wave Mott insulator ($d$-Mott), which is {\it not} adiabatically continuable to a band insulator; by doping the $d$-Mott state with holes, depending on the magnitude of $U$, it gives way to a $d$-wave superconducting state, a two-flavor spin-1/2 Fermi liquid (FL), or a spin-3/2 FL.     
\end{abstract}
\maketitle

The phase diagram of weakly correlated metals tends to be relatively simple---superconductivity can occur at low temperatures induced by weak attractive interactions and spin-density waves (SDW) and/or charge density waves (CDW) occur under special circumstances when the Fermi surface is sufficiently well nested.  However,  with strong interactions, there is no reason {\em not to}  have multiple ordered phases. Inverting this logic, one might expect competing phases to be a {\it generic} feature of strongly interacting systems. So it becomes increasingly important to find a {\it simple and solvable} model of strong interactions, which could serve as a paradigmatic example showing these complexities of competing orders. 

\begin{figure}[b]
\includegraphics[scale=0.21]{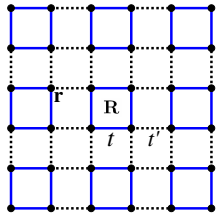}
\caption{(Color online) The schematic representation of the checkerboard Hubbard model. The hopping amplitudes are $t=1$ on the solid bonds (blue) and $t^\prime\ll 1$ on the dashed bonds (black). $\br$ labels sites and $\bR$ the plaquettes. The lattice spacing between nearest neighboring sites is set to 1 for simplicity.}
\label{fig_lattice}
\end{figure}
\begin{figure}[b]
\includegraphics[scale=0.36]{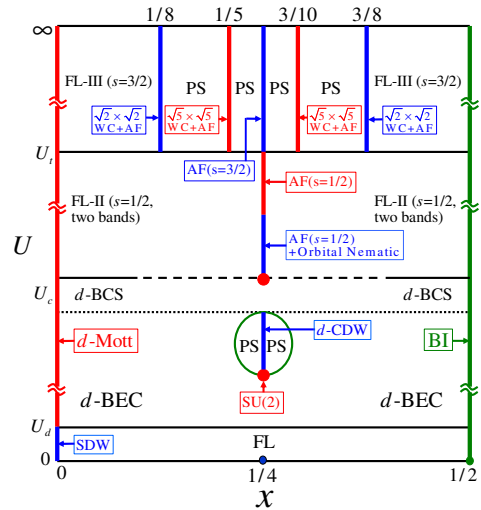}
\caption{(Color online) Phase diagram of the checkerboard Hubbard model for $0\leq x\leq 1/2$ and all $U>0$.  
Abbreviations:  ``FL''=Fermi liquid; ``$s$=$n/2$'' = spin-$n/2$;  ``PS'' = phase separation; ``AF'' = antiferromagnet; ``WC'' = Wigner crystal;  ``$d$-BCS'' = $d$-wave superconductor;  ``$d$-Mott,'' ``$d$-BEC,'' and ``$d$-CDW'' are phases made of $d$-wave two-particle bound-states (hard-core bosons) which are, respectively, a  Bose-Mott insulator, a superfluid, and a charge density wave; ``SDW''= spin-density wave; and ``BI''= band insulator.  The various phases are described in the text.}
\label{fig_phase_diagram_1}
\end{figure}

The Hubbard model is the simplest model of a strongly interacting electron gas, but alas, no well-controlled solution exists in more than one dimension. Here we study the Hubbard model, Eq.~(\ref{Hubbard}), on a checkerboard lattice with hopping matrix element $t$ between nearest-neighbor sites on elementary square plaquettes and $t^\prime$ between sites on neighboring plaquettes.  (See Fig.~\ref{fig_lattice}.) For $t^\prime=t$, this model reduces to the usual (still unsolved) Hubbard model on a square lattice.~\cite{Assa02} 
For $t^\prime \ll t$, where it is a crystal of weakly coupled ``Hubbard clusters,'' we are able to establish a number of features of the zero temperature phase diagram, even for strong interactions, $U/t > 1$, using $t^\prime/t$ as a small parameter.  (See Fig.~\ref{fig_phase_diagram_1}.) Particularly striking is the large number of zero temperature phases;  we have established the existence of at least 16 distinct phases and there are undoubtedly more in the portions of the phase diagram for which we have not yet obtained a solution. For instance, at $x=0$ and $U>{\cal O}(t^\prime)$, the ground state is a novel $d$-wave Mott insulator ($d$-Mott), which is a {\it true} new state of matter. More generally, the details of the phase diagram depend sensitively on the choice of clusters. For example, the phase diagram of the dimerized Hubbard model,\cite{Tsai06} largely consists of a single Fermi liquid (FL) phase, plus a band-insulating (BI) and spin-1/2 antiferromagnetic (AF) insulating phase. However, multiple competing phases appear to be a common feature of the strong interacting limit. 
 
Recently, there have been a number of cluster-dynamical mean-field theory (DMFT) studies of correlated electrons.  The complexity of the phase diagram of the present model in the $t^\prime \ll t$ limit, and the strong dependence on the precise type of clusters raise questions concerning the validity of this approach. Conversely, solvable cluster models in the small $t^\prime$ limit can serve as interesting benchmark tests for such approximate approaches, and for future analog simulations with cold fermionic atoms in optical lattices.

{\it Model Hamiltonian}. The checkerboard Hubbard model, originally studied in Ref.~\onlinecite{Tsai06}, has Hamiltonian
\begin{eqnarray}\label{Hubbard}
{\cal H} =-\sum_{\bra\br\br^\prime\ket,\sigma}t_{\br\br^\prime}c^\dag_{\br\sigma} c_{\br^\prime\sigma} + \frac U 2\sum_{\br}  [\hat n_\br-1]^2,
\end{eqnarray}
where $c^\dag_{\br\sigma}$ creates an electron on site $\br$ with spin polarization $\sigma=\uparrow,\downarrow$ and $\hat n_\br=\sum_{\sigma}c^\dag_{\br\sigma}c_{\br\sigma}$. Here $t_{\br\br^\prime}$ is the hopping matrix element from site $\br^\prime$ to $\br$ and $\bra\br\br^\prime\ket$ denotes nearest-neighbor sites. $t_{\br\brprime} = t$ or $t_{\br\brprime} =t^\prime\ll t$ when $\bra\br\br^\prime\ket$ are a pair of sites connected, respectively, by a solid bond or a dashed bond shown in Fig.~\ref{fig_lattice}. We set $t=1$ as our energy units. Note that the model with uniform on-site repulsion $U$ preserves the point group symmetry, $C_{4v}$, of the square lattice. The density of electrons per site is defined to be $n_{el} \equiv 1-x$ where $x$ is the density of ``doped holes" per site.  Since the model is particle-hole symmetric, we restrict our discussion to $0\le x \le 1$. 

Treating $t^\prime$ as a small parameter permits us to solve 
this model using perturbation theory. In the unperturbed $t^\prime=0$ problem, the 2D lattice consists of decoupled four-site square plaquettes. 
The Hubbard model of a four-site plaquette is exactly solvable. The eigenstates of the decoupled 2D system are direct products of the eigenstates on each plaquette. For most densities, $x$, the unperturbed ground-state is degenerate, so we use degenerate (or near-degenerate) perturbation theory to derive an effective Hamiltonian in the low energy state space. 

{\it Ground states of an isolated plaquette}. The eigenstates of a single plaquette can be specified by the number of doped holes, $Q_h$, the total and $z$ component of the spin, and the familiar orbital labels ``$s$" (even under $C_4$---i.e., $90^\circ$ rotation), ``$p_x\pm ip_y$" (phase changed by $\pm\pi/2$ under $C_4$), and ``$d$" (odd under $C_4$). 
For a single plaquette, the $Q_h=0$ ground state is unique  
for any positive $U$ and has $d$-wave symmetry. In the $Q_h=1$ sector, the plaquette ground state for $U<U_t\approx 18.6$ has spin 1/2 and  $p_x\pm ip_y$ orbital symmetry, i.e., it is four-fold degenerate
corresponding to spin polarizations $s=\pm 1/2$ and orbital ``chiralities'' $\tau=\pm 1/2$. However, for $Q_h=1$ and $U>U_t$, the ground state is spin 3/2 and orbital $s$ wave, so it is still four-fold degenerate.
When there are two holes ($Q_h=2$), the ground state is unique and has $s$-wave symmetry for all $U$.  
The $Q_h=3$ ground state is (trivially) spin 1/2 and $s$ wave.

In adding holes to the system, the issue arises whether it is energetically cheaper to add two holes to one plaquette or one hole to each of two plaquettes.  This is determined by the sign of the pair binding energy\cite{White92} $\Delta\equiv E_0(0)+E_0(2)-2E_0(1)$,
where $E_0(Q_h)$ is the ground state energy of one plaquette with $Q_h$ holes. When $U<U_c\approx 4.6$, $\Delta$ is negative which indicates that doped two holes prefer to stay in the same plaquette, effectively forming a hard-core boson. When $U>U_c$, $\Delta$ is positive; i.e., two holes repel each other. Note, in Fig.~\ref{fig_phase_diagram_1}, that the critical values of $U$ at which level crossings occur for the isolated plaquette figure prominently in the phase diagram of the perturbed system, as well.

{\it Effective Hamiltonians}. Starting from these states, for various ranges of $x$ and $U$, we can derive the effective low energy Hamiltonian in powers of $t^\prime$. Although this procedure reduces the number of dynamical degrees of freedom, it still leaves us with a non-trivial many-body problem, which is only solvable in certain cases;  the phases exhibited in Fig.~\ref{fig_phase_diagram_1} are those whose existence we have established, but there are compelling reasons to expect additional phases to exist in the portions of the $x$-$U$ plane that we have not fully analyzed.  We will provide more details of the analysis, and a discussion of the regions of the phase diagram that have only been partially analyzed in a future publication.\cite{long}
 
For $0 < U \ll {\cal O}({t^\prime})$, the interactions are weak, so the zeroth order description is in terms of bands (see below), and except at $x=0$ [where the Fermi surface (FS) is nested] and $x=1/2$, where there is a BI, we expect a FL description to be valid.

For $0< x < 1/2$ and ${\cal O}(\sqrt{t^\prime}) < U <  U_c - {\cal O}(t^\prime)$,  $2x$ of the plaquettes are occupied by a pair of holes, and $(1-2x)$ have no holes.  Identifying hole pairs as hard-core bosons, the effective Hamiltonian is 
\begin{eqnarray}
\label{bosons}
H^{(1)} =-t^{(1)}\sum_{\bra\bR\bRprime\ket} b^\dag_\bR b_\bRprime+V^{(1)} \sum_{\bra\bR\bRprime\ket} \rho_\bR\rho_\bRprime,
\end{eqnarray}
where the bosonic creation operator $b_\bR^\dagger$ creates a hole pair on plaquette $\bR$, $\rho_\bR=b^\dag_\bR b_\bR$ is the number operator, and there is an implicit no-double occupancy (hard-core) constraint.  Because the zero-hole state has $d$-wave symmetry and the two-hole state has $s$-wave symmetry, $b_\bR$ is a charge $2e$ field which transforms like a $d$-wave under $C_{4}$.\cite{Scalapino96} Here $t^{(1)}$ is the effective hopping of bosons and $V^{(1)}$ the repulsion between nearest neighbor bosons, both of order  ${t^\prime}^2$. Their explicit dependences on $U$ are somewhat complicated, but we have computed them exactly.\cite{long}

For $0< x < 1/2$  and $U_c - {\cal O}(t^\prime) < U < U_c + {\cal O}(t^\prime)$, both singly charge and doubly charged plaquettes occur, so the problem maps onto a rather complicated version of the  Boson-Fermion model, as discussed in Ref.~\onlinecite{Tsai06}. 

For $0< x \le 1/4$ and $U_c + {\cal O}(t^\prime) < U < U_t$, the low energy states are a mixture of no-hole and one-hole plaquettes, where the one-hole states are further distinguished by two possible total-spin polarizations $s=\pm 1/2$ and two orbital chiralities $\tau=\pm 1/2$---i.e., $p_x\pm ip_y$. Consequently, the effective Hamiltonian is a two-flavor version of the $t$-$J$-$V$ model:
\begin{eqnarray}\label{effective_Hamiltonian_2}
H^{(2)} =-t^{(2)}\sum_{\bra\bR\bRprime\ket,s,\tau} \phi_{\bR,\bR^\prime}f^\dag_{\bR,s,\tau}f_{\bR^\prime,s,-{\tau}}+H^{(2,2)},
\end{eqnarray}
where $f^\dag_{\bR,s,\tau}$ creates a fermion with spin polarization $s=\pm 1/2$ and chirality $\tau=\pm 1/2$, and there is a no-double occupancy constraint which we have left implicit. Here $\phi_{\bR,\bR^\prime}$ is $+1$ ($-1$) if the effective bond $\bR\bR^\prime$ is along the $\hat x$ ($\hat y$) direction. The hopping parameter $t^{(2)}$ is order of $t^\prime$, while $H^{(2,2)}$ refers to terms  of order ${t^\prime}^2$:
\begin{eqnarray}\label{exchange2}
&&H^{(2,2)}=
J^{(2)}\sum_{\bra\bR\bRprime\ket}\bS_\bR\cdot\bS_\bRprime 
+V^{(2)}\sum_{\bra\bR\bRprime\ket} n_\bR n_\bRprime \\\nonumber
&&\qquad+\sum_{\bra\bR\bRprime\ket}
\big[{\cal J}_x\tau^x_\bR\tau^x_\bRprime
+{\cal J}_y\tau^y_\bR\tau^y_\bRprime 
+{\cal J}_z\tau^z_\bR\tau^z_\bRprime\big],\\\nonumber
&&~~~+\sum_{\bra\bR\bRprime\ket} \bS_\bR\cdot\bS_\bRprime 
\big[{\cal J}^\prime_x\tau^x_\bR\tau^x_\bRprime
+{\cal J}^\prime_y\tau^y_\bR\tau^y_\bRprime 
+{\cal J}^\prime_z\tau^z_\bR\tau^z_\bRprime\big],
\end{eqnarray}
where $\bS_\bR$, $\btau_\bR$, and $n_\bR$ are spin, pesudo-spin and density operators on plaquette $\bR$ respectively. Strictly speaking, there are additional ``pair-hopping'' terms, which we have computed but do not display;  for $x=1/4$, where $H^{(2,2)}$ is the leading term in the effective Hamiltonian, the pair-hopping terms vanish.

For $U>U_t$, the one-hole ground state of a single plaquette has spin-3/2
so the effective Hamiltonian for $0<x\le 1/4$ is a $t$-$J$-$V$ model for spin-3/2 fermions 
\begin{eqnarray}\label{effective_Hamiltonian_3}
&&H^{(3)}=-t^{(3)}\sum_{\bra\bR\bRprime\ket,s} f^\dag_{\bR s}f_{\bRprime s} \\\nonumber
&&\qquad~~~+J^{(3)}\sum_{\bra\bR\bRprime\ket}\bS_\bR\cdot\bS_\bRprime
+V^{(3)}\sum_{\bra\bR\bRprime\ket} n_\bR n_\bRprime,
\end{eqnarray}
where $f^\dag_{\bR s}$ is the plaquette fermion creation operator on plaquette $\bR$ with spin polarization $s=\pm 1/2,\pm 3/2$ (as always there is no double occupancy allowed), and $\bS_\bR$ and $n_\bR$ are corresponding spin and density operators on plaquette $\bR$. In this case, the effective hopping $t^{(3)}$ is order of ${t^\prime}^3$ while $J^{(3)}$ and $V^{(3)}$ are order of ${t^\prime}^2$.
Consequently, this model always occurs in what, for the spin-1/2 model, is considered an unphysical limit $J^{(3)}$, $V^{(3)} \gg t^{(3)}$. 

For $U>U_c + {\cal O}(t^\prime)$ and $1/4 < x < 1/2$, the effective Hamiltonian is, again, of the form presented in Eqs.~(\ref{effective_Hamiltonian_2}) and~(\ref{exchange2}) (for $U < U_t$) or Eq.~(\ref{effective_Hamiltonian_3}) (for $U > U_t$), but with different values of the couplings.  Moreover, whereas for $0 \le x \le 1/4$ the vacuum state is identified with the $x=0$ state with 4 electrons per plaquette, so the mean density of fermions per plaquette is $4x$; for $1/4 \le x \le 1/2$, the vacuum state has 2 electrons per plaquette and the mean density of fermions is $(2-4x)$.

When $x>1/2$, the low energy degrees of freedom are always plaquettes fermions, each of which is just an ordinary electron. The effective Hamiltonian is the $t$-$J$-$V$ model, having the same form as Eq.~(\ref{effective_Hamiltonian_3}), but for spin-1/2 fermions with different effective parameters $t^{(4)}$, $V^{(4)}$, and $J^{(4)}$. Here, $t^{(4)}$ is order of $t^\prime$, while $J^{(4)}$ and $V^{(4)}$ are order of ${t^\prime}^2$. So $t^{(4)}\gg V^{(4)}$, $J^{(4)}$.   

{\it Phase diagram}. Much of the structure of the phase diagram is obvious from the effective Hamiltonian.  We  now sketch some less obvious aspects of the analysis.

{\it Zero doping}. Because the zero-hole ground state of a single plaquette is unique and there is a finite gap, at $x=0$ the unperturbed ground state in the limit $t^\prime\to 0$ is a direct product state and small $t^\prime$ produces only perturbative corrections which do not change any qualitative properties of the ground state.
This is an insulating phase with no broken symmetry.  However, despite the fact that there are four electrons per unit cell, this state is not adiabatically connected to a BI state, since it transforms according to a non-trivial representation of the point group:  from the $d$-wave character of the single-plaquette wave function, it follows that the many-body wave function changes sign under $90^\circ$ rotation about a plaquette center.  One can think of this as a Mott insulating state with one $d$-wave boson per plaquette; hence, we call it a ``$d$-Mott'' state.    In terms of macroscopic observable properties, this phase has at least two identifying features: (i) The pair-field pair-field correlation function $\langle c^\dag_{{\bf 0}\uparrow} c^\dag_{\br\downarrow} c_{{\bR}\downarrow} c_{\bR+\brprime^\uparrow}\rangle$, although it falls exponentially with distance $|{\bf R}|$, has an asymptotic $d$-wave symmetry (for large $|{\bf R}|$) upon $90^\circ$ rotation of $\br$ or $\br^\prime$ separately. (ii) It is an orbital paramagnet.\cite{Moreo90,long} This $d$-Mott phase is a {\it genuine} new state of matter, in contrast with the similar state in ladder systems \cite{Lin98,Wu03} where there is no $C_4$ symmetry to unambiguously distinguish it from a band insulator.

The fact that the $d$-Mott phase is not adiabatically related to a BI implies that, even for $x=0$, there must be a phase transition as a function of decreasing $U$. For fixed, small $t^\prime$, when $U$ gets small enough, the gap in the isolated plaquette is no longer large compared to $t^\prime$.  Specifically, when $U\ll t^\prime$ the kinetic energy is dominant and the $U$ term can be treated through a weak-coupling Hartree-Fock approximation. Since there are four sites per unit cell, there are for $U=0$ four bands as follows:
\begin{eqnarray}
\label{bands}
\epsilon_\bk\!=\!\pm\sqrt{(t-t^\prime)^2+4tt^\prime\cos^2{k_x}}
\pm\!\sqrt{(t-t^\prime)^2+4tt^\prime\cos^2{k_y}}.\nonumber
\end{eqnarray}
The top and bottom bands are well separated from the two middle bands $\epsilon_{\bk,\pm}\approx\pm 2t^\prime (\cos^2{k_x}-\cos^2{k_y})$ by a gap of approximately $2t$. 
Particle-hole symmetry fixes the Fermi energy at 0 for $x=0$, so that the FS coincides with the lines $\cos{k_x}=\pm\cos{k_y}$ where the two bands touch. Consequently, the FS is perfectly nested, and any weak positive $U$ induces an SDW ground state ordering at $(\pi/2,\pi/2)$, in which the FS is fully gapped.

{\it For $x=1/2$} there is an insulating ground-state which is smoothly connected to the BI state at $U=0$.  For $x=1$, there are no electrons, which is trivially an insulating state.

{\it The hard-core boson model} in Eq.~(\ref{bosons}) has been studied~\cite{Hebert01} extensively numerically, and its $T=0$ phase diagram is known. For most $x\in(0,1/2)$ (i.e., for boson concentration between 0 and 1), it has a uniform superfluid phase, which inherits the $d$-wave symmetry of the bosons, but has no nodal quasiparticles;  this is labeled $d$-BEC in Fig.~\ref{fig_phase_diagram_1}.  At $x=1/4$, the boson density is 1/2 per plaquette;  in this case, it is easy to see (by mapping the problem to an equivalent XXZ model) that, for $V^{(1)}/t^{(1)} >2$, the ground state is a $\sqrt{2} \times \sqrt{2}$ CDW state of $d$-wave bosons, while for $V^{(1)}/t^{(1)} < 2$, the ground state is superfluid. At the critical point separating these two phases, $V^{(1)}/t^{(1)}  = 2$, the effective Hamiltonian has an emergent $SU(2)$ symmetry. In the present case, we find that at $U=U_c$ and $U=U_s\approx 2.7$, $V^{(1)}/t^{(1)} =2$, and that $V^{(1)}/t^{(1)} >2$ for $U_s < U < U_c$ and $V^{(1)}/t^{(1)} <2$ for $U <U_s$. Around the $d$-CDW line in the phase diagram shown in Fig.~\ref{fig_phase_diagram_1}, there is a small two-phase coexistence or phase separation (PS) region because the transition from the $d$-CDW state to the $d$-BEC is first-order.\cite{Batrouni00,Hebert01}

{\it The boson-fermion model} which applies for $U \sim U_c$ and $0 < x < 1/2$ is quite complicated, and has not been fully analyzed.  However, as was shown in~Ref.~\onlinecite{Tsai06}, it can be analyzed for $x\gtrsim 0$ taking advantage of the fact that the fermions and bosons are dilute.  A similar analysis applies in terms of the particle-hole transformed model for $x\lesssim 1/2$. 
By increasing $U$, there exists a crossover (the dotted line in Fig.~\ref{fig_phase_diagram_1}) from the $d$-BEC phase to a BCS-like superconducting phase ($d$-BCS without nodal quasiparticles). 
For even larger $U$, a phase transition into a spin-1/2 FL phase with two flavors (bands) of fermions (FL-II) is expected for the effective model in Eq.~(\ref{effective_Hamiltonian_2}) at small $x$.

{\it The spin-3/2 $t$-$J$-$V$ model}, Eq.~(\ref{effective_Hamiltonian_3}), is similar to the spin-1/2 model in the weak hopping limit;  this model was studied in detail.\cite{Kivelson90,Emery90} The ground state depends on the ratio of $J^{(3)}/V^{(3)}$. In the present problem, we obtain values of $V^{(3)}$ an order of magnitude larger than $J^{(3)}$. Thus, the ground state is a spin-3/2 FL (FL-III) for $x<1/8$. For $x=1/8$ (one fermion on every second plaquette), the system forms a $\sqrt{2} \times\! \sqrt{2}$ Wigner crystalline (WC) state, on top of which the residual antiferromagnetic interactions  induce $2\times 2$ AF order ($\sqrt{2} \times \! \sqrt{2}$  WC + AF). 
For $x = 1/4$, every plaquette is occupied by a single fermion, whose spins order to yield a $\sqrt{2} \times\! \sqrt{2}$ AF.  When $x=1/5$, there is a concentration $4(1/4 - 1/5) = 1/5$ of unoccupied plaquettes which order in a $\sqrt{5}\times\! \sqrt{5}$ WC in the background of the $\sqrt{2} \times\! \sqrt{2}$ antiferromagnetic order ($\sqrt{5} \times \! \sqrt{5}$  WC + AF). A particle-hole transformation results in a second copy of the same sequence of phases (with somewhat different energetics) for $1/4 < x < 1/2$.

{\it For $U_c<U<U_t$ and $x=1/4$}, there is one fermion per plaquette, so the only terms in the effective Hamiltonian that operate are those in $H^{(2,2)}$. This is a complex model with a spin and pseudospin on each plaquette.  We have solved it by approximating the ground-state as a direct product of spin and pseudospin factors.  Since the spin interactions are AF and isotropic, they form the well-understood N\'eel ground state of the spin-1/2 AF, in which $\bra \bS_\bR\cdot\bS_\bRprime\ket\approx -1/3$ for nearest-neighboring $\bR$ and $\bR^\prime$. Then, the effective psueduo-spin Hamiltonian is 
\begin{eqnarray}
&&H^{(2,2)}_{\mathrm{pseudo}} =\sum_{\bra\bR\bRprime\ket}\sum_{\alpha=x,y,z} 
\bar{{\cal J}}_\alpha\tau^\alpha_\bR\tau^\alpha_\bRprime,
\end{eqnarray}
where $\bar{{\cal J}}_\alpha={\cal J}_\alpha+{\cal J}^\prime_\alpha\bra\bS_\bR\cdot\bS_\bRprime\ket$. The ordering of the pseudo-spins is determined by the $\bar{{\cal J}}_\alpha$ with largest absolute value. When $U<U_n\approx 7.3$, $-\bar{{\cal J}}_x>-\bar{{\cal J}}_y>\bar{{\cal J}}_z>0$, which indicates that the ground state of pseudospins is ferromagnetically ordered along the $\hat x$ direction. This orbital ordering corresponds to the fact that the electron density spontaneously breaks the $C_4$ rotational symmetry to $C_2$ with no breaking of translational symmetry, so this is an ``electron nematic'' or `` orbital nematic'' phase.\cite{Kivelson98} While for $U_n<U<U_t$, there is no such nematic ordering.\cite{long}

{\it For $x=1/2$}, with two electrons per plaquette, the insulating ground state is adiabatically connected to the BI state at $U=0$.  Since the plaquette fermion hopping is the dominant term, for $1/2<x<3/4$ and $3/4<x<1$, the ground state [except, probably, for a narrow region $|x-3/4| < {\cal O}(t^\prime)$] is a spin-1/2 FL, while at $x=3/4$, the ground state is a spin-1/2 AF.

{\it Finite temperature}. The finite $T$ phase diagram is also interesting\cite{Kocharian06} and worth future study. For instance, at $x=1/4$ and for $U_c<U<U_n$, the $T=0$ phase is a spin-1/2 AF and orbital nematic. At any finite $T$, the spin order is lost, leaving a pure (Ising) nematic phase up to a nonzero critical temperature.
 
We thank D. J. Scalapino for helpful suggestions.  This work was supported, in part, by the NSF DMR-0531196 (S.A.K.) and the DOE DE-FG02-06ER46287 (H.Y. and W.T.). 


\end{document}